\documentclass[twocolumn]{revtex4}
\usepackage{epsfig}
\usepackage{amssymb}
\usepackage{amsmath}
\usepackage{amsfonts}

\newcommand{\R}{\mathbb{R}}
\newcommand{\C}{\mathbb{C}}

\newcommand{\fa}{\mathfrak{a}}

\newcommand{\fz}{\mathfrak{z}}

\newcommand{\fD}{\mathfrak{D}}

\newcommand{\be}{\begin{equation}}
\newcommand{\ee}{\end{equation}}
\newcommand{\bea}{\begin{eqnarray}}
\newcommand{\eea}{\end{eqnarray}}
\newcommand{\nn}{\nonumber}
\newcommand{\kt}{\rangle}
\newcommand{\br}{\langle}

\newcommand{\ed}{\end{document}}

\newcommand{\pbr}{\prec\!}
\newcommand{\pkt}{\!\succ}

\newcommand{\bi}{\begin{itemize}}
\newcommand{\ei}{\end{itemize}}
\newcommand{\One}{{\stackrel{\leftrightarrow}{1}}}
\newcommand{\Ep}{{\stackrel{\leftrightarrow}{\mbox{\large{$\varepsilon$}}}}}
\newcommand{\MU}{{\stackrel{\leftrightarrow}{\mbox{\large{$\mu$}}}}}
\newcommand{\Mu}{{\stackrel{\;\leftrightarrow_\prime}{\mbox{\large{$\mu$}}}}}

\begin{document}
\title{Propagation of Electromagnetic Waves in Linear Media and
Pseudo-Hermiticity}

\author{A.~Mostafazadeh$^1$ and F.~Loran$^2$}
\address{$^1$Department of Mathematics, Ko\c{c} University, Sariyer 34450,
Istanbul, Turkey\\
$^2$Department of Physics, Isfahan University of Technology,
Isfahan, Iran.}

\begin{abstract}
We express the electromagnetic field propagating in an arbitrary
time-independent non-dispersive medium in terms of an operator
that turns out to be pseudo-Hermitian for Hermitian dielectric and
magnetic permeability tensors $\Ep$ and $\MU$. We exploit this
property to determine the propagating field. In particular, we
obtain an explicit expression for a planar field in an isotropic
medium with $\Ep=\varepsilon\One$ and $\MU=\mu\One$ varying along
the direction of the propagation. We also study the scattering of
plane waves due to a localized inhomogeneity.

\hspace{7.5cm}{Pacs numbers: 03.50.De, 41.20.Jb, 42.25.Bs}
\end{abstract}

\maketitle

The study of electromagnetic (EM) fields propagating in a linear
(non-dispersive) medium is one of the oldest problems of physics.
Most textbook treatments of this problem begin with the assumption
of harmonic time-dependence. In this letter, we present a
systematic solution of this problem that does not rely on this
assumption and instead makes use of the notion of
``pseudo-Hermitian operator'' that was originally developed in
\cite{p1,p23} to deal with ${\cal PT}$-symmetric Hamiltonians
\cite{bender-1998}.\footnote{For other applications of
pseudo-Hermitian operators see \cite{other}.}

Consider the propagation of EM fields in a linear source-free
medium with time-independent dielectric and inverse magnetic
permeability tensors $\Ep=\Ep(\vec x)$ and $\Mu=\Mu(\vec x)$,
\cite{jackson}. Then the electric and magnetic fields, $\vec E$
and $\vec B$, satisfy Maxwell's equations:
    \bea
    &&\vec\nabla\cdot\vec D=0,~~~~~\vec\nabla\cdot\vec B=0,
    \label{max-1}\\
    &&\dot{\vec B}+\fD\vec E=0,~~~~~
    \dot{\vec D}-\fD \vec H=0,
    \label{max-2}
    \eea
where $\vec D:=\Ep\,\vec E$, $\vec H:=\Mu\vec B$, a dot means a
time-derivative, and $\fD$ denotes the curl operator; $\fD\vec
F:=\vec\nabla\times\vec F$ for any vector field $\vec F$. Our aim
is to solve (\ref{max-2}) for $\vec E=\vec E(x,t)$ and $\vec
B=\vec B(\vec x,t)$ in terms of the initial fields $\vec E_0:=
\vec E(\vec x,0)$ and $\vec B_0:=\vec B(\vec x,0)$. We will
consider lossless media for which $\Ep(\vec x)$ and $\Mu(\vec x)$
are (Hermitian) positive-definite matrices for all $\vec
x\in\R^3$.

We begin our study by expressing Eqs.~(\ref{max-2}) in terms of
$\vec E$ and $\vec B$. Taking the time-derivative of the second of
these equations and using the result in the first, we obtain
$\ddot{\vec E}+\Omega^2 {\vec E}=0$ where
    \be
    \Omega^2:=\Ep^{-1}\fD\Mu\fD.
    \label{Omega}
    \ee
We can easily solve $\ddot{\vec E}+\Omega^2 {\vec E}=0$ to find
    \be
    \vec E(\vec x,t)=\cos(\Omega t)\vec E_0(\vec x)+
    \Omega^{-1}\sin(\Omega t)\dot{\vec E}_0(\vec x),
    \label{E=}
    \ee
where $\dot{\vec E}_0(\vec x):=\dot{\vec E}(\vec
x,0)=\Ep^{-1}\fD\Mu\vec B_0(\vec x)$ and
    \bea
    &&\cos(\Omega t):=\sum_{n=0}^\infty
    \frac{(-1)^n}{(2n)!}\;(t^2\Omega^2)^n,
    \label{cos}\\
    &&\Omega^{-1}\sin(\Omega t):=t\sum_{n=0}^\infty
    \frac{(-1)^n}{(2n+1)!}\;(t^2\Omega^2)^n.
    \label{sin}
    \eea
In view of the first equation in (\ref{max-2}), the magnetic field
is given by $\vec B(\vec x,t)=\vec B_0(\vec x)-\int^t_0ds\,\fD\vec
E (\vec x,s)$, \footnote{Eliminating $\vec E$ in (\ref{max-2})
yields $\ddot{\vec B}+\Gamma^2\vec B=0$ where
$\Gamma^2:=\fD\Ep^{-1}\fD\Mu$. The solution of $\ddot{\vec
B}+\Gamma^2\vec B=0$ is $\vec B=\cos(t\Gamma)\vec B_0+
\Gamma^{-1}\sin(t\Gamma)\dot{\vec B}_0$. Applying $\Ep^{-1}\fD\Mu$
to both sides of this equation and using (\ref{max-2}) and the
identity $\Ep^{-1}\fD\Mu\Gamma^2=\Omega^2\Ep^{-1}\fD\Mu$, we
obtain $\dot{\vec E}=-\Omega\sin(\Omega t)\vec E_0+\cos(\Omega
t)\dot{\vec E}_0, $ which is consistent with (\ref{E=}).}.

Relation (\ref{E=}) is of limited practical importance, because in
general its right-hand side involves infinite derivative
expansions. We can choose the initial fields such that they are
eliminated by a positive integer power of $\Omega^2$. This leads
to an infinite class of exact solutions of Maxwell's equations
with polynomial time-dependence. In order to use (\ref{E=}) in
dealing with physically more interesting situations, we will
express (\ref{cos}) and (\ref{sin}) as integral operators and
compute the corresponding integral kernels
(propagators).\footnote{One might try to compute the terms in the
derivative expansions appearing on the right-hand side of
(\ref{E=}). Even for the simplest choices for the medium and the
initial data, this yields a power series solution of Maxwell's
equation in time whose summation proves to be extremely difficult
if not impossible. Approximating the terms in this series and
summing the corresponding approximate series seems to be also a
formidable task.} This requires a closer look at $\Omega^2$.

Let $\pbr\vec F,\vec G\pkt:=\int_{\R^3}dx^3\vec F(\vec x)^*\cdot
\vec G(\vec x)$, where $\vec F$ and $\vec G$ are vector fields.
Then ${\cal H}:=\{\vec F:\R^3\to\C^3|\pbr\vec F,\vec
F\pkt\,<\!\infty\}$ together with the inner product $\pbr\cdot,
\cdot\pkt$ form a Hilbert space. It is easy to check that the curl
operator $\fD$ is actually a Hermitian operator acting in ${\cal
H}$. That is $\pbr\vec F,\fD\vec G\pkt=\pbr\fD\vec F,\vec G\pkt$.
The same is true about $\fD\Mu\fD$. But $\Omega^2$ is not
Hermitian. Its adjoint ${\Omega^2}^\dagger$, which is defined by
the condition $\pbr\vec F,\Omega^2\vec
G\pkt=\pbr{\Omega^2}^\dagger\vec F,\vec G\pkt$, satisfies
${\Omega^2}^\dagger=\Ep{\Omega^2}\Ep^{-1}$. This means that
$\Omega^2:{\cal H}\to{\cal H}$ is an $\Ep$-pseudo-Hermitian
operator \cite{p1}. Here we view $\Ep$ as an operator acting in
${\cal H}$ according to $(\Ep \vec F)(\vec x):=\Ep(\vec x)\vec
F(\vec x)$. Indeed, because $\Ep(\vec x)$ is a positive-definite
matrix for all $\vec x$, the operator $\Ep$ is a positive-definite
(metric) operator. This in turn implies that it defines a new
positive-definite inner product that renders $\Omega^2$
self-adjoint \cite{p1}; letting $\pbr\vec F,\vec G
\pkt_{\varepsilon}:=\pbr\vec F,\Ep\vec G\pkt$, we find $\pbr\vec
F,\Omega^2\vec G \pkt_{\varepsilon}=\pbr\Omega^2\vec F,\vec G
\pkt_{\varepsilon}$. Furthermore, $\Omega^2$ may be mapped to a
Hermitian operator $h:{\cal H}\to{\cal H}$ via a similarity
transforms \cite{p23}. A possible choice for $h$ is
\cite{jpa-2003-2004b}
    \be
    h:=\Ep^{\frac{1}{2}}\Omega^2\Ep^{-\frac{1}{2}}=
    \Ep^{-\frac{1}{2}}\fD\Mu\fD\Ep^{-\frac{1}{2}}.
    \label{h=}
    \ee
Note that because $\Ep(\vec x)$ and $\Mu(\vec x)$ are assumed to
be positive matrices for all $\vec x$, they have a unique positive
square root \cite{reed-simon}. This in turn implies that $h$ is a
positive operator with a nonnegative spectrum and a unique
positive square root $h^{\frac{1}{2}}$, \footnote{This is because
$h=\fa^\dagger \fa$ where
$\fa:=\Mu^{\frac{1}{2}}\fD\Ep^{-\frac{1}{2}}$.}.

Because $h$ is Hermitian, we can use its spectral resolution to
compute any function $\Phi$ of $h$, \cite{messiah}. In light of
$\Phi(\Omega^2)=\Ep^{-\frac{1}{2}}\Phi(h)\,\Ep^{\frac{1}{2}}$,
this allows for the calculation of the action of $\Phi(\Omega^2)$
on any vector field $\vec G$:
    \be
    \Phi(\Omega^2)\vec G(\vec x)=\int_{\R^3}\!\!\!dy^3\,
    \Ep(\vec x)^{-\frac{1}{2}}\,\br\vec x|\Phi(h)|\vec y\kt\,
    \Ep(\vec y)^{\frac{1}{2}}\vec G(\vec y),
    \label{fG}
    \ee
where we have used Dirac's bra-ket notation.

To demonstrate the effectiveness of the above method we consider
the textbook \cite{yeh} problem of the planar propagation of the
initial fields
    \be
    \vec E_0(z)={\cal E}_0(z)\;e^{-i k_0 z}\, \hat i,~~~~
    \vec B_0(z)={\cal B}_0(z)\;e^{-i k_0 z}\, \hat j,
    \label{e1}
    \ee
along the $z$-axis in an isotropic medium with $\Ep(\vec
x)=\varepsilon(z)\One$ and $\Mu(\vec x)=\mu(z)^{-1}\One$, where
$\vec x=:(x,y,z)$, $\One$ is the identity matrix, ${\cal E}_0$ and
${\cal B}_0$ are given envelope functions, $k_0$ is the principal
wave number at which the Fourier transform of $\vec E_0(z)$ and
$\vec B_0(z)$ are picked, $\hat i$ and $\hat j$ are the unit
vectors along the $x$- and $y$-axes, and $\varepsilon(z)$ and
$\mu(z)$ are respectively (the $z$-dependent) dielectric and
magnetic permeability constants. We will in particular consider
the cases that $\varepsilon(z)$ and $\mu(z)$ tend to constant
values as $z\to\pm\infty$.

For this configuration all the fields are independent of $x$ and
$y$-coordinates and we have
$\Omega^2=\varepsilon(z)^{-1}p\,\mu(z)^{-1}p$, where
$p:=-i\frac{d}{dz}$,
    \be
    h=\varepsilon(z)^{-\frac{1}{2}}p\;\mu(z)^{-1}p
    \;\varepsilon(z)^{-\frac{1}{2}},
    \label{sh=}
    \ee
and $\dot{\vec E}_0(z):=\varepsilon(z)^{-1}\dot{\vec D}_0(z)=
\varepsilon(z)^{-1}\fD\vec B_0(z)=\varepsilon(z)^{-1} [ik_0 {\cal
B}_0(z)-{\cal B}'_0(z)]\; e^{-ik_0z}\,\hat i.$

In order to determine the spectral resolution of $h$ we need to
solve the time-independent Schr\"odinger equation for the
position-dependent-mass Hamiltonian (\ref{sh=}), i.e.,
    \be
    -\varepsilon(z)^{-\frac{1}{2}}\,\frac{d}{dz}\left(
    \mu(z)^{-1}\frac{d}{dz}\,
    [\varepsilon(z)^{-\frac{1}{2}}\psi(z)]\right)=\omega^2\psi(z).
    \label{ode}
    \ee
Because of the above-mentioned asymptotic behavior of
$\varepsilon(z)$ and $\mu(z)$, the eigenfunctions of $h$ are the
solutions of (\ref{ode}) fulfilling the bounded boundary
conditions at $\pm\infty$. Also note that $\omega^2\in\R^+$,
because $h$ is a positive operator.

For an arbitrary $\varepsilon$ we cannot solve (\ref{ode})
exactly. Therefore, we employ the WKB approximation. To do this we
express $\psi$ in its polar representation: $\psi=R~e^{iS}$ where
$R$ and $S$ are real-valued functions. Inserting $\psi=R~e^{iS}$
in (\ref{ode}) gives
    \bea
    &&S'(z)^2+Q(z)=
    \omega^2\varepsilon(z)\,\mu(z),
    \label{wkb1}\\
    &&\frac{d}{dz}\,[\mu(z)^{-1}R_-(z)^2S'(z)]=0,
    \label{wkb2}
    \eea
where
    \be
    Q(z):=-\frac{[\mu(z)^{-1}R'_-(z)]'}{\mu(z)^{-1}R_-(z)},~~~~
    R_-(z):=\frac{R(z)}{\varepsilon(z)^{\frac{1}{2}}}.
    \label{Q=}
    \ee
WKB approximation amounts to neglecting $Q(z)$ in (\ref{wkb1}).
This yields
    \be
    S(z)=\omega\: u(z)+c_1,~~~~~
    R_-(z)=c_2\:\varepsilon(z)^{-\frac{1}{4}}\mu(z)^{\frac{1}{4}},
    \label{wkb4}
    \ee
where
    \be
    u(z):=\int_0^z d\fz~v(\fz)^{-1},~~~~
    v(z):=[\varepsilon(z)\mu(z)]^{-\frac{1}{2}},
    \label{u=}
    \ee
and $c_1,c_2$ are possibly $\omega$-dependent integration
constants. Using these choices for $S$ and $R_-$ and fixing $c_1$
and $c_2$ appropriately, we find the following $\delta$-function
normalized eigenfunctions for all $\omega\in\R$.
    \be
    \psi_{\omega}(z):=
    \frac{e^{i\omega u(z)}}{\sqrt{2\pi\,v(z)}} .
    \label{psi=}
    \ee

Next, we use $\psi_\omega$ to express $\cos(h^{\frac{1}{2}}t)$ and
$h^{-\frac{1}{2}}\sin(h^{\frac{1}{2}}t)$ in terms of their
spectral resolution:
    \bea
    &&\cos(h^{\frac{1}{2}}t)=\int_{-\infty}^\infty d\omega~
    \cos(\omega t)~|\psi_\omega\kt\br\psi_\omega|,
    \label{e15-1}\\
    &&h^{-\frac{1}{2}}\sin(h^{\frac{1}{2}}t)=
    \int_{-\infty}^\infty d\omega~
    \frac{\sin(\omega t)}{\omega}~|\psi_\omega\kt\br\psi_\omega|.
    \label{e15}
    \eea
Using (\ref{psi=}) -- (\ref{e15}) and the identities
$\int_{-\infty}^\infty d\omega\; e^{ia\omega}=2\pi\delta(a)$ and
$\int_{-\infty}^\infty d\omega\; e^{ia\omega}/\omega=\pi i\; {\rm
sign}(a)$, with $\delta$ denoting the Dirac delta function and
${\rm sign}(x):=x/|x|$ for $x\neq 0$ and ${\rm sign}(0):=0$, we
find
    \bea
    \br z|\cos(h^{\frac{1}{2}}t)|w\kt&=&
    \int_{-\infty}^\infty d\omega~\cos(\omega t)~
    \psi_\omega(z)\,\psi_\omega(w)^*\nn\\
    &=&\frac{1}{2}\;[v(z)v(w)]^{-\frac{1}{2}}
    \Delta(z,w;t),
    \label{Kc}\\
    \br z|h^{-\frac{1}{2}}
    \sin(h^{\frac{1}{2}}t)|w\kt&=&
    \int_{-\infty}^\infty\!\! d\omega\,\omega^{-1}\sin(\omega t)\,
    \psi_\omega(z)\,\psi_\omega(w)^*\nn\\
    &=&\frac{1}{4}\;[v(z)v(w)]^{-\frac{1}{2}}\Sigma(z,w;t),
    \label{Ks}
    \eea
where
$$\Delta(z,w;t):=\delta[u(w)-u(z)+t]+\delta[u(w)-u(z)-t],~~~~~~$$
$$\Sigma(z,w;t):={\rm sign}[u(w)-u(z)+t]-{\rm sign}[u(w)-u(z)-t].$$

Because $u$ is a monotonically increasing function that vanishes
only at $z=0$, it is invertible and its inverse $u^{-1}$ is also a
monotonically increasing function with a single zero at $z=0$.
This implies that the quantity
     \be
    w_\pm(z,t):=u^{-1}(u(z)\pm t)
    \label{wpm}
    \ee
is the only zero of $u(w)-(u(z)\pm t)$. Hence,
    $$\delta[u(w)-u(z)\pm t]=
    \frac{\delta[w-w_\mp(z,t)}{|u'(w_\mp(z,t))|}.$$
In view of this relation and (\ref{u=}), we have
    \be
    \Delta(z,w;t)=\frac{\delta[w-w_-(z,t)]}{v(w_-(z,t))^{-1}}
    +\frac{\delta[w-w_+(z,t)]}{v(w_+(z,t))^{-1}}.
    \label{e21}
    \ee
Furthermore, because both $u$ and $u^{-1}$ are monotonically
increasing, for $t>0$ we have $w_-(z,t)<w_+(z,t)$ and
    \be
    \Sigma(z,w;t)=
    \left\{\begin{array}{cc}
    2 &~~{\rm for}~~w_-(z,t)<w< w_+(z,t),\\
    0 & {\rm otherwise.}\end{array}\right.
    \label{e22}
    \ee

Next, we compute the action of $\cos(\Omega t)$ on an arbitrary
test function $f(z)$. To do this we set
$\Phi(\Omega^2)=\cos(\Omega t)$ in (\ref{fG}) and use (\ref{Kc})
and (\ref{e21}) to evaluate the corresponding integral. This
yields
    \be
    \cos(\Omega t)f(z)=f_-(z,t)+f_+(z,t),
    \label{e23}
    \ee
where
    $$ f_\pm(z,t):=\frac{1}{2}\left[\frac{\varepsilon(w_\pm(z,t))
    v(w_\pm(z,t))}{\varepsilon(z)
    v(z)}\right]^{\frac{1}{2}}f(w_\pm(z,t)).$$
Similarly, setting $\Phi(\Omega^2)=\Omega^{-1}\sin(\Omega t)$ in
(\ref{fG}) and using (\ref{Ks}) and (\ref{e22}) we find
    \be
    \Omega^{-1}\sin(\Omega t)f(z)=
    \frac{[\varepsilon(z)v(z)]^{-\frac{1}{2}}}{2}\!
    \int_{w_-(z,\tau)}^{w_+(z,\tau)}\!\!\!\!\!\!
    dw\left[\frac{\varepsilon(w)^{\frac{1}{2}}
    f(w)}{v(w)^{\frac{1}{2}}}\right].
    \label{e24}
    \ee
Finally, we use (\ref{u=}), (\ref{e23}) and (\ref{e24}) to express
(\ref{E=}) as
    \bea
    \vec
    E(z,t)&=&\frac{1}{2}
    \left[\frac{\mu(z)}{\varepsilon(z)}\right]^{\frac{1}{4}}
    \left\{\left[\frac{\varepsilon(w_-(z,t))}{\mu(w_-(z,t))}
    \right]^{\frac{1}{4}}\right.\!\!
    \vec E_0(w_-(z,t))+\nn\\
    &&\vspace{.5cm}\left[\frac{\varepsilon(w_+(z,t))}{\mu(w_+(z,t))}
    \right]^{\frac{1}{4}}\!\!
    \vec E_0(w_+(z,t))+\nn\\
    &&\vspace{.5cm}\left.\int_{w_-(z,t)}^{w_+(z,t)} dw~
    \mu(w)^{\frac{1}{4}}\varepsilon(w)^{\frac{3}{4}}
    \dot{\vec E}_0(w)\right\}.
    \label{e25}
    \eea
{According to (\ref{wpm}),  $w_\pm(z,0)=z$. This shows that
setting $t=0$, the integral on the right-hand side of (\ref{e25})
disappears and the remaining terms add up to give $\vec E_0(z)$.
Similarly we can use (\ref{u=}) and (\ref{wpm}) to establish $\dot
w_\pm(z,0)=\pm v(z)=\pm[\varepsilon(z)\mu(z)]^{-\frac{1}{2}}$ and
use the latter to show that setting $t=0$ in the time-derivative
of the right-hand side of (\ref{e25}) yields $\dot{\vec E}_0(z)$.

In vacuum where $v=(\varepsilon\mu)^{-1/2}=c$, WKB approximation
is exact, $u(z)=z/c$, $w_\pm(z,t)=z\pm ct$, and (\ref{e25}) gives
    \bea
    \vec E(z,t)&=&\frac{1}{2}
    \left\{\vec E_0(z-ct)+\vec E_0(z+ct)+\right.\nn\\
    &&\hspace{.5cm}\left.\frac{1}{c}\int_{z-ct}^{z+ct} dw~
    \dot{\vec E}_0(w)\right\},
    \eea
which is precisely D'Alembert's solution of the 1+1 dimensional
wave equation \cite{Strauss}.}

In general, (\ref{e25}) is a valid solution of Maxwell's
equations, if WKB approximation is reliable. This is the case
whenever $|Q(z)|$ is negligibly smaller than the right-hand side
of (\ref{wkb1}), \footnote{If
$\varepsilon(z)=a\mu(z)[b+\int_0^zd\fz\:\mu(\fz)^{-1}]^{-4}$ for
some constants $a$ and $b$, $Q(z)=0$ and WKB approximation is
exact.}. In view of (\ref{wkb4}) and (\ref{u=}) this condition
takes the form
    \be
    \frac{v^2}{2}\left|\frac{vv''-\frac{1}{2}\,{v'}^2}{{v}^2}+
    \frac{\mu\mu''-\frac{3}{2}\,{\mu'}^2}{{\mu}^2}\right|\ll \omega^2,
    \label{e26}
    \ee
where we have suppressed the $z$ dependence of $v$ and $\mu$. Due
to the asymptotic behavior of $\mu$ and $\varepsilon$, $v$ tends
to constant values as $z\to\pm\infty$. This in turn implies that
the square root of the left-hand side of (\ref{e26}) has a least
upper bound that we denote by $\omega_{\rm min}$. In this case,
(\ref{e26}) means $|\omega|\gg \omega_{\rm min}$. Recalling the
role of $\omega$ in our derivation of (\ref{e25}), we can view
this condition as a restriction on the choice of the initial
conditions. More specifically, (\ref{e25}) is a good approximation
provided that for all $\omega\in[-\omega_{\rm min},\omega_{\rm
min}]$, $\br\psi_\omega|\varepsilon^{\frac{1}{2}}|\vec
E_0(z)|\kt\approx 0$ and
$\br\psi_\omega|\varepsilon^{\frac{1}{2}}|\dot{\vec
E}_0(z)|\kt\approx 0$. For a planar laser pulse with initial
envelope functions ${\cal E}_0$ and ${\cal B}_0$ picked far away
from the region where $\varepsilon$ and $\mu$ vary significantly,
these conditions hold for $c^{-1}\omega_{\rm min} \ll |k_0|$. The
same is true for an initial plane wave with sufficiently large
wave number $|k_0|$.

{Next, we wish to elaborate on the relationship between our
approach and the standard application of the WKB approximation in
solving Maxwell equations, particularly for the effectively
one-dimensional model we have been considering. The first step in
the direct application of the WKB approximation to Maxwell
equations is to assume a harmonic time-dependence for the
solution. This yields an ordinary differential equation that can
be treated using the WKB approximation, \cite{yeh}. If we apply
$\varepsilon^{-1/2}$ to eigenfunctions (\ref{psi=}) of the
Hermitian operator $h$ we obtain (WKB-approximate) eigenfunctions
of the pseudo-Hermitian operator $\Omega^2$,
\cite{jpa-2003-2004b}. It is very easy to check that these are
solutions of the Maxwell equations (\ref{max-2}) with a harmonic
time-dependence. Up to a trivial normalization constant, they
coincide with the conventional WKB solutions:
    $$
    \vec{E}_\omega(z,t)=\left[\frac{\mu(z)}{\varepsilon(z)}
    \right]^{\frac{1}{4}} e^{i\omega t} e^{i\omega u(z)}.$$
Furthermore, if we make the additional assumption that $\vec
E_\omega$ form a complete set in the sense that for any solution
$\vec E(z,t)$ of Maxwell's equations, there are $c(\omega)\in\C$
such that
    \be
    \vec E(z,t)=\int_{-\infty}^\infty d\omega\:
    c(\omega)\vec E_\omega(z,t),
    \label{wkb-old}
    \ee
then we can use the initial electric and magnetic field to
determine $c(\omega)$ and use the result to evaluate the integral
in (\ref{wkb-old}). The end result of this lengthy calculation is
identical to (\ref{e25}).

The observations that $\Omega^2$ is pseudo-Hermitian and that it
can be mapped by a similarity transformation to a Hermitian
operator $h$ provide a simple justification for the
above-mentioned assumptions regarding harmonic time-dependence and
completeness of the conventional WKB solutions. The consequences
of the pseudo-Hermiticity of $\Omega^2$ is not limited to the
application of the WKB approximation. For instance, it implies the
existence of exact solutions of Maxwell's equations that have
harmonic time-dependence and form a complete set.}

As a concrete example of the application of (\ref{e25}), suppose
that $\varepsilon$ has a Lorentzian shape and $\mu$ is a constant:
    \be
    \varepsilon(z)=\varepsilon_0
    \left[1+a\left(1+\gamma^{-2}z^2\right)^{-1}\right],~~~~~~~~~
    \mu=\mu_0,
    \label{epsilon=}
    \ee
where $\varepsilon_0,a,\gamma,\mu_0$ are positive constants and
$\varepsilon_0\mu_0=c^{-2}$. Then, by inspection, we can show that
    $$\omega_{\rm min}\leq \frac{c\sqrt{3a(1+\nu a)}}{2\gamma},$$
where $\nu:=1+45/256\approx 1.176$. For $a<1$, $\omega_{\rm
min}<c\gamma^{-1}$. This means that WKB approximation and
consequently (\ref{e25}) are valid provided that the allowed
$\omega$ values be much larger than $c\gamma^{-1}$.

In order to implement (\ref{e25}), we must compute $w_\pm$. This
involves the evaluation and inversion of $u$. For the choice
(\ref{epsilon=}), we expand $u$ and $u^{-1}$ in power series in
the inhomogeneity parameter $a$ and perform a perturbative
calculation of $w_\pm$. This yields
    \bea
    w_\pm(z,t) &=& z\pm ct \mp a\left\{
    \frac{\gamma\theta_\pm(z,t)}{2}\right\}\pm a^2\Big\{
    \frac{\gamma}{16}\times\nn\\
    &&
    \left[\lambda(z,t)\,\theta_\pm(z,t)+\nu_\pm(z,t)\right]\Big\}+
    {\cal O}(a^3),~~~~
    \label{wpm=}
    \eea
where
    \bea
    \theta_\pm(z,t)&:=&\tan^{-1}\left(\frac{\gamma ct}{
    \gamma^2+z(z\pm ct)}\right),\nn\\
    \lambda(z,t)&:=&1+\frac{4\gamma^2}{\gamma^2+(z\pm ct)^2},\nn\\
    \nu_\pm(z,t)&:=&\frac{\gamma ct[\gamma^2-z(z\pm ct)]}{(
    \gamma^2+z^2)[\gamma^2+(z\pm ct)^2]}.\nn
    \eea
In view of (\ref{epsilon=}) and (\ref{wpm=}), we also have
    \bea
    \left[\frac{\varepsilon(w_\pm(z,t))}{
    \varepsilon(z)}\right]^{\frac{1}{4}}&=&
    1\mp a~\xi(z,t)\pm a^2\Big\{\xi(z,t)+\nn\\
    &&\zeta(z,t)\mp\frac{3}{2}\,\xi(z,t)^2\Big\}
    +{\cal O}(a^3),~~~~
    \label{ep-ep}
    \eea
where
    \bea
    \xi(z,t)&:=&\frac{\gamma^2\, ct(2z\pm
    ct)}{4(\gamma^2+z^2)[\gamma^2+(z\pm ct)^2]},\nn\\
    \zeta(z,t)&:=&
    \frac{\gamma^3(z\pm ct)}{4[\gamma^2+(z\pm ct)^2]^2}.\nn
    \eea
With the help of (\ref{wpm=}) and (\ref{ep-ep}) we can use
(\ref{e25}) to determine the dynamical behavior of the EM fields
for initial configurations of the form (\ref{e1}) provided that
the initial fields do not violate the condition of the reliability
of the WKB approximation and that we can neglect the third and
higher order contributions in powers of $a$. For example we can
use this method to determine the effect of the inhomogeneity
(\ref{epsilon=}) on the planar propagation of a Gaussian laser
pulse \footnote{We have studied a pulse with ${\cal
E}_0(z)=A~e^{-\frac{(z+L)^2}{2\sigma^2}}$, ${\cal B}_0(z)=0$,
$A,L,\sigma\in\R^+$, $L\gg\gamma$, and $L\gg\sigma$. Then the WKB
approximation is valid whenever $\gamma^{-1}\ll k_0$. We do not
report the results here for lack of space.}.

Another application of our results is in the solution of the
scattering problem. It is not difficult to see that for
$t\to\infty$ (i.e., $ct\gg z,\gamma$),
    \bea
    && w_\pm(z,t)\to \left\{\begin{array}{ccc}
    z\pm ct+\Delta r(z)&{\rm for}&z\neq 0\\
    \pm\, ct\pm\Delta r(0)&{\rm for}&z=0\end{array}\right.,
    \label{w-limit}\\
    && \left[\frac{\varepsilon(w_\pm(z,t))}{
    \varepsilon(z)}\right]^{\!\!\frac{1}{4}}\to
    \Delta\rho(z),
    \label{ep-limit}
    \eea
where
    \bea
    \Delta r(z)&:=&-\frac{a\gamma }{2}\tan^{-1}
    \left(\frac{\gamma}{z}\right)+\nn\\
    &&\frac{a^2\gamma}{16}\big[\tan^{-1}
    \left(\frac{\gamma}{z}\right)-\frac{\gamma z}{\gamma^2+z^2}\big]+
    {\cal O}(a^3),\nn\\
    \Delta\rho(z)&:=&\left[\frac{\varepsilon_0}{
    \varepsilon(z)}\right]^{\frac{1}{4}}+{\cal
    O}(a^3)\nn\\
    &=&1-\frac{a\gamma^2}{4(\gamma^2+z^2)}+
    \frac{5a^2\gamma^4}{32(\gamma^2+z^2)^2}+{\cal O}(a^3).\nn
    \eea

\begin{figure}
\epsffile{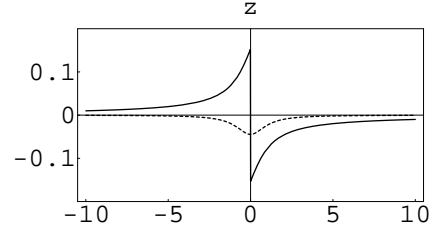} \caption{Plots of $\Delta r$ (full curve)
and $\Delta\rho$ (dotted curve) as a function of $z$ for $a=0.2$
in units where $\gamma=1$.\label{fig1}}
\end{figure}

According to (\ref{e25}), (\ref{w-limit}) and (\ref{ep-limit}),
the scattering of an initial plane wave, with $\vec E_0=e^{-ik_0
z}\hat i$ and $\vec B_0=\vec 0$, by the inhomogeneity
(\ref{epsilon=}) results in a change in the amplitude and phase
angle of the wave that are respectively given by $\Delta\rho(z)$
and $-k_0\Delta r(z)$. Specifically, as $t\to\infty$,
    \be
    \vec E(z,t)\to \vec E_s(z,t):=
    \vec E(z,t)\Big|_{a=0}\; \Delta\rho(z)\,
    e^{-ik_0\Delta r(z)},
    \label{scatter-1}
    \ee
where
    $\vec E(z,t)\big|_{a=0}:=\frac{1}{2}\left(e^{-ik_0 (z-ct)}+
    e^{-ik_0 (z+ct)}\right)\hat i.$
The predictions of (\ref{scatter-1}) should be experimentally
verifiable in typical interferometry experiments \footnote{One can
imagine using a Mach-Zender interferometer to detect the effect of
the inhomogeneity in the interference pattern of two beams one
travelling through the medium and the other through the vacuum.}.
Figure~\ref{fig1} shows the plots of $\Delta\rho$ and $\Delta r$
for $a=0.2$. As seen from this figure $\Delta r$ has a
discontinuity at $z=0$.

\begin{figure}
\epsffile{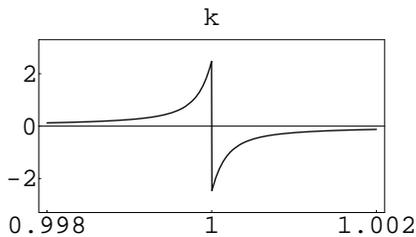} \caption{Plot of $\Delta\tilde E(k_0,k)$
as a function of $k$ for $\gamma=10^{-3} m$ and $k_0=10^7 m^{-1}$.
$k$ is measured in units of $k_0=10^7 m^{-1}$. \label{fig2}}
\end{figure}

A quantity of direct physical relevance is the Fourier transform
$\vec{\tilde E}_s(k,t):=\frac{1}{2\pi}\int_{-\infty}^\infty dz\,
e^{ikz} \vec{E}_s(z,t)$ of $\vec{E}_s(z,t)$. Up to linear terms in
$a$, it is given by
    $$\vec{\tilde
    E}_s(k,t)=\left[\delta(k_0-k)+a\Delta\tilde E(k_0,k)+{\cal
    O}(a^2)\right]\cos(k_0ct)\hat i,$$
where
    $\Delta\tilde
    E(k_0,k):=\gamma\left[\frac{1-e^{-\gamma|k_0-k|}}{4(1-k/k_0)}-
    \frac{e^{-\gamma|k_0-k|}}{8}\right].$
Figure~\ref{fig2} shows the graph of $\Delta\tilde E(k_0,k)$ as a
function of $k$. Again there is a discontinuity at $k=k_0$.

To summarize, we have obtained a closed form expression (\ref{E=})
for the propagation of EM fields in an arbitrary non-dispersive
stationary medium that yields the fields in terms of a
pseudo-Hermitician operator $\Omega^2$. This allows for
formulating the problem in terms of an equivalent Hermitian
operator $h$. Using the spectral resolution of $h$ and the WKB
approximation we have found an explicit formula, namely
(\ref{e25}), for the propagating fields and demonstrated its
application for the scattering of plane waves moving in an
inhomogeneous non-dispersive medium. Although similar spectral
techniques have previously been used in dealing with EM waves
\cite{spectral}, we believe that our treatment provides a more
straightforward and systematic solution for this problem. Indeed
to our knowledge an expression for the propagating field as
general and explicit as Eq.~(\ref{e25}) has not appeared in the
literature previously. Another important aspect of our method is
the wide range of its applications, e.g., it can be used to study
wave propagation in inhomogeneous fibers, observation of
superfluid vortices, etc.

Our results may be generalized in various directions. For example,
for the cases that $\Ep$ fails to be Hermitian, one may appeal to
the notion of weak pseudo-Hermiticity \cite{weak} and use the
results of \cite{jmp-06b} to obtain an appropriate equivalent
Hermitian operator $h$ to $\Omega^2$. One may also incorporate the
dispersion effects by letting the $\Ep$ and $\Mu$ that appear in
the eigenvalue equation for $h$ to depend (via a dispersion
relation) on $\omega$. This will lead to a modification of
(\ref{e25}) that we plan to explore in future.

\acknowledgments We wish to thank \"Ozg\"ur
M\"ustecapl{\scriptsize l}o\u{g}lu and Alphan Sennaro\u{g}lu for
helpful discussions.

\ed